# Exponential parameterization of the neutrino mixing matrix — comparative analysis with different data sets and CP violation

K. Zhukovsky[1], A. Borisov[2]

**Abstract**

The exponential parameterization of Pontecorvo-Maki-Nakagawa-Sakata mixing matrix for neutrino is used for comparative analysis of different neutrino mixing data. The $\mathbf{U}_{PMNS}$ matrix is considered as the element of the SU(3) group and the second order matrix polynomial is constructed for it. The inverse problem of constructing the logarithm of the mixing matrix is addressed. In this way the standard parameterization is related to the exponential parameterization exactly. The exponential form allows easy factorization and separate analysis of the rotation and the CP violation. With the most recent experimental data on the neutrino mixing (May 2016), we calculate the values of the exponential parameterization matrix for neutrinos with account for the CP violation. The complementarity hypothesis for quarks and neutrinos is demonstrated to hold, despite significant change in the neutrino mixing data. The values of the entries of the exponential mixing matrix are evaluated with account for the actual degree of the CP violation in neutrino mixing and without it. Various factorizations of the CP violating term are investigated in the framework of the exponential parameterization.

[1] Deptartment of Theoretical Physics, Faculty of Physics, M.V.Lomonosov Moscow State University, Moscow 119991, Russia. Phone: +7(495)9393177, e-mail: zhukovsk@physics.msu.ru

[2] Deptartment of Theoretical Physics, Faculty of Physics, M.V.Lomonosov Moscow State University, Moscow 119991, Russia. Phone: +7(495)9393177. e-mail: borisov@phys.msu.ru

## 1. Introduction

The Standard Model (SM) [1]–[3] gives the description of electromagnetic and weak interactions by the unified theory. The neutrino plays important role in it. The original formulation of the SM presumed the neutrino had zero mass. However, the existence of at least three massive neutrino states, $\nu_1$, $\nu_2$, $\nu_3$, was proposed and, consequently, the neutrino oscillations [4] were predicted by Pontecorvo [5], [6]. The discovery of the neutrino oscillations was awarded the Nobel Prize in physics in 2015. The neutrino has three flavours and the latter vary during the neutrino propagation. The neutrino states constitute the full and normalized orthogonal basis, confirmed by numerous experiments and observations of neutrino oscillations with solar, atmospheric, reactor and accelerator neutrinos [7], [8], [9]. The neutrino flavour states, $\nu_e$, $\nu_\mu$, $\nu_\tau$, are constructed of different mass states, $\nu_1$, $\nu_2$, $\nu_3$, by the unitary Pontecorvo-Maki-Nakagawa-Sakata (PMNS) matrix $\mathbf{U}_{PMNS}$ [10]:

$$|\nu_\alpha\rangle = \sum_{i=1,2,3} \mathbf{U}^*_{PMNS\,\alpha i} |\nu_i\rangle, \quad \mathbf{U}_{PMNS\,\alpha i} \equiv \langle \nu_\alpha | \nu_i \rangle, \qquad (1)$$

similarly to the way it is done for quarks by the CKM matrix. Mixing in the lepton sector of the SM means that a charged $\mathbf{W}$-boson interacts with mass states of charged leptons $e$, $\mu$, $\tau$ and with neutrino states $\nu_1$, $\nu_2$, $\nu_3$. The boson $\mathbf{W}^+$ decays into a pair of lepton $\alpha$ and neutrino $i$ with the amplitude $\mathbf{U}_{\alpha i}$. The above formula (1) evidences that the production of the pair of the lepton $\alpha$ and of the neutrino in the state $\alpha$ implies the superposition of all three neutrino mass states, $\nu_1$, $\nu_2$, $\nu_3$. There are several proposals of the mixing matrix parameterization, as well as there are different parameterizations for quarks. This, however, does not cause any contradiction, if the unitarity, which is the only strict requirement, is ensured. The most common standard parameterization $\mathbf{U}_{st}$ for three neutrino species is implemented by the unitary 3×3 mixing matrix $\mathbf{U}_{st}$:

$$\mathbf{U}_{PMNS} = \mathbf{U}_{st} \mathbf{P}_{Mj}, \qquad (2)$$

where

$$\mathbf{U}_{st} = \begin{pmatrix} c_{12}c_{13} & s_{12}c_{13} & s_{13}e^{-i\delta_{CP}} \\ -s_{12}c_{23} - c_{12}s_{23}s_{13}e^{i\delta_{CP}} & c_{12}c_{23} - s_{12}s_{23}s_{13}e^{i\delta_{CP}} & s_{23}c_{13} \\ s_{12}s_{23} - c_{12}c_{23}s_{13}e^{i\delta_{CP}} & -c_{12}s_{23} - s_{12}c_{23}s_{13}e^{i\delta_{CP}} & c_{23}c_{13} \end{pmatrix}, \qquad (3)$$

$$\mathbf{P}_{\mathrm{Mj}} = diag\left(1, e^{i\alpha_1/2}, e^{i\alpha_2/2}\right), \tag{4}$$

$c_{ij} = \cos\theta_{ij}$, $s_{ij} = \sin\theta_{ij}$, $i,j$=1,2,3, and $\mathbf{P}_{\mathrm{Mj}}$ stands for the possible Majorana nature of the neutrino. For Majorana neutrinos, identical to their antiparticles, the phases $\alpha_{1,2} \neq 0$ play role in the processes with violation of the lepton number. The sterile neutrino, which does not interact with **Z-** and with **W-**bosons (see, for example, [11], [12], [13]), is also possible, but not considered here. The role of the matrix $\mathbf{U}_{st}$ in the parameterization (3) is very similar to that of the CKM matrix in quark mixing [14], [16]–[19], and the form of the matrix (3) is identical to that of the standard CKM mixing matrix for quarks. Historically first proposal of the mixing matrix parameterisation for quarks by Kobayashi and Maskawa differed in phase placement from (3):

$$\mathbf{V}_{\mathrm{KM}} = \begin{pmatrix} c_1 & -s_1 c_3 & -s_1 s_3 \\ s_1 c_2 & c_1 c_2 c_3 - s_2 s_3 e^{i\delta_{CP}} & c_1 c_2 s_3 + s_2 c_3 e^{i\delta_{CP}} \\ s_1 s_2 & c_1 s_2 c_3 + c_2 s_3 e^{i\delta_{CP}} & c_1 s_2 s_3 - c_2 c_3 e^{i\delta_{CP}} \end{pmatrix}, \tag{5}$$

$s_i = \sin\theta_i$ $c_i = \cos(\theta_i)$, $i, j$=1,2,3. When $\theta_2 = \theta_3 = 0$ we obtain the Cabibbo form of the mixing matrix in quark sector, where $\theta_1 = \theta_c$ is the Cabibbo angle. In the standard parameterization the Cabibbo case is realized when $\theta_{23} = \theta_{13} = 0$ and $\theta_{12} = \theta_c$. Moreover, the small parameter for quark mixing exists: $\lambda = \sin\theta_c \approx 0.22$ [20], which is not present for neutrinos. While parameterisations of the mixing matrix may differ from each other — physics does not depend on its choice — we are free to choose the most convenient for us. The PMNS matrix is fully determined by four parameters: three mixing angles $\theta_{12}$, $\theta_{23}$, $\theta_{13}$ and the phase $\delta$ in charge of the CP violation [14].

Other parameterisations of the neutrino mixing matrix exist (see, for example, [21]–[28]), of which the exactly unitary tri-bimaximal parameterization (TBM) (see, for example, [21]) of $\mathbf{U}_{\mathrm{PMNS}}$ was for long rather consistent with the experimental data. Completed parameterization, based on the TBM pattern, was described in [21], [28], [29]). The TBM parameterization has the mixing angles $\theta_{12} = \arctan\left(1/\sqrt{2}\right) \cong 35.25°$, $\theta_{23} = \pi/4 = 45°$, which agree very well with the values, obtained from experimental sets, and only $\theta_{13} = 0$ contradicts recent data, which indicates it is not zero. The TBM mixing matrix reads as follows:

$$\mathbf{U}_{TBM} = \begin{pmatrix} \sqrt{2/3} & 1/\sqrt{3} & 0 \\ -1/\sqrt{6} & 1/\sqrt{3} & -1/\sqrt{2} \\ -1/\sqrt{6} & 1/\sqrt{3} & 1/\sqrt{2} \end{pmatrix}. \tag{6}$$

Since there are no convincing reasons for the TBM form to be exact, and, moreover, it follows from recent experimental data that $\theta_{13} \neq 0$, the approximate parameterisations of the PMNS matrix are developed, based upon the deviations from the TBM form (see, for example, [29], [30]). In contrast with the parameterisations in the quark sector, constructed with a single parameter, parameterisations for the neutrino mixing include 3 parameters, defining the deviations of the reactor, solar and atmospheric neutrino mixing angles from their tri-bimaximal values. Triminimal expansion around the bimaximal basis for quark and lepton mixing parameterization matrices was developed in [30]. The authors also discussed the unified description between different kinds of parameterizations for quark and lepton sectors: the standard parameterizations, the Wolfenstein-like parameterizations and the triminimal parameterizations in the context of the quark-lepton complementarity (QLC) hypothesis [22], [31]. The latter consists in the phenomenological relations of quark and lepton mixing angles $\theta_{qij}$ and $\theta_{ij}$ in the standard parameterization: $\theta_{12} + \theta_{q12} = 45°$, $\theta_{23} + \theta_{q23} = 45°$. The QLC is an important subject of this study, and there are many other studies in this line, such as [32], [33], [34]. In what follows we will explore this topic in the context of the rotation axes direction in three dimensional space in the exponential parameterization of the mixing matrix.

The pioneering proposal of the unitary exponential parameterization for the neutrino mixing was done in [35] by A. Strumia, F. Vissani. The exponential parameterization for quarks was proposed in [37]; very similar parameterization for neutrinos was studied in [36] in the following form:

$$\mathbf{U}_{\exp} = \exp \mathbf{A}, \tag{7}$$

where

$$\mathbf{A} = \mathbf{A}_0 = \begin{pmatrix} 0 & \lambda_1 & \lambda_3 e^{i\delta} \\ -\lambda_1 & 0 & -\lambda_2 \\ -\lambda_3 e^{-i\delta} & \lambda_2 & 0 \end{pmatrix}. \tag{8}$$

The anti-Hermitian form of the matrix **A** ensures the unitarity of the transforms by the mixing matrix $\mathbf{U}_{\exp}$ (7) (see [38]). The parameter $\delta$ accounts for the CP violation and the parameters $\lambda_i$ are responsible for the flavour mixing. For neutrinos, in contrast with that for quarks, the mixing angles $\theta_{12}$ and $\theta_{23}$ are large and, therefore, the hierarchy in the exponential quark mixing matrix, based on the single parameter $\lambda$: $\lambda_1 \propto \lambda$, $\lambda_2 \propto \lambda^2$, $\lambda_3 \propto \lambda^3$, does not hold for neutrinos. For $\delta=2\pi n$ and for $\delta=\pi(2n+1)$ the matrix $\mathbf{A}_0$ (8) turns into the three dimensional rotation matrix in angle–axis presentation [39]. The most important advantage of the exponential parameterization of the mixing matrix respectively to the commonly known standard parameterization is that the exponential parameterization allows easy factorization of the rotational part, the CP-violating terms and possible Majorana term [39], [40]. Note, that the above exponential parameterization with the matrix $\mathbf{A}_0$ (8) is not the only one possible, and it just represents the simplest attempt to account for the mixing and for the CP violation in the framework of the most general exponential parameterization. *Importantly, the exponential matrix $\mathbf{U}_{\exp}$ (7) with the ansatz* (8) *does not reduce to the standard parameterization $\mathbf{U}_{st}$ (3)*. The difference in the results is negligible for small values of $\delta$, but it becomes significant for big values of $\delta$. In the following chapters we will address this topic in details.

## 2. Exponential parameterization and the matrix logarithm

In general, an exponential of a matrix $\hat{A}$ can be treated similarly to the exponential of the operator if viewed as the expansion in series $e^{\hat{A}} = \sum_{n=0}^{\infty} \hat{A}^n / n!$; the latter can be computed with any given precision, if proper number of terms are calculated. It can be reduced to the second order of $\hat{A}$ matrix polynomial with the help of the Cayley-Hamilton theorem as done in [40]. While this algebraic method gives explicit analytical expression for the exponential matrix in terms of the zero, first and second orders of the exponential, the calculations are bulky. Recently, this problem was reinvestigated in [41] in the context of the fundamental representation of the SU3 group. We recall that $\mathbf{U}_{PMNS}$ belongs the SU3 group and, omitting the details of [41], we pick up the main useful for us result, i.e. that for any SU(3) group element, generated by a traceless 3×3 Hermitian matrix **H,** the following representation holds [41]:

$$\exp[i\theta\mathbf{H}] = \sum_{k=0,1,2}\left[\mathbf{H}^2 + \frac{2}{\sqrt{3}}\mathbf{H}\sin(\phi+2\pi k/3) - \frac{1}{3}\mathbf{I}(1+2\cos(2(\phi+2\pi k/3)))\right]\times$$
$$\frac{\exp\left(\frac{2}{\sqrt{3}}i\theta\sin(\phi+2\pi k/3)\right)}{1-2\cos(2(\phi+2\pi k/3))}, \qquad (9)$$

where the scale for the $\theta$ parameter space is set by the common normalization

$$\text{tr}[\mathbf{H}^2] = 2. \qquad (10)$$

The above formula (9) with the help of the Laplace transforms can be written as the ordinary differential equation (DE) [41]:

$$\exp[i\theta\mathbf{H}] = \left(\mathbf{H}^2 - i\mathbf{H}\frac{d}{d\theta} - \mathbf{I}\left(1+\frac{d^2}{d^2\theta}\right)\right)\sum_{k=0,1,2}\frac{\exp\left(\frac{2}{\sqrt{3}}i\theta\sin(\phi+2\pi k/3)\right)}{1-2\cos(2(\phi+2\pi k/3))}, \qquad (11)$$

giving link to the operational approach for DE [42]–[48]. So expressed as a matrix polynomial, the group element depends on the group rotation angle $\theta$ and on the sole invariant $\det(\mathbf{H})$, which is encoded cyclometrically as another angle (see [41]):

$$\phi = \frac{1}{3}\left(\arccos\left(\frac{3}{2}\sqrt{3}\det(\mathbf{H})\right) - \frac{\pi}{2}\right). \qquad (12)$$

Upon distinguishing in the exponential parameterization $\mathbf{U}_{\text{exp}} = \exp\mathbf{A}$ (7) the $i\theta$ factor to match the l.h.s. of (9) and with account for the normalization (10) we obtain

$$\theta = \left(-\text{tr}\left[\mathbf{A}^2/2\right]\right)^{1/2} \qquad (13)$$

for the $\theta$ parameter, which is in essence the rotation angle, describing the displacement from the SU(3) group origin. Now with the help of the formula (9) we can express the $\mathbf{U}_{\text{PMNS}}$ matrix, being a group element for the fundamental representation of SU(3), as a second order matrix polynomial of a Hermitian generating matrix $\mathbf{H}$ with coefficients (12), consisting of elementary trigonometric functions of the sole invariant $\det(\mathbf{H})$ (see (12)). In what follows we will apply this technique to the best fit neutrino mixing matrix.

Now let us study the inverse problem, which in essence consists in finding the logarithm of the $\mathbf{U}_{\text{PMNS}}$ matrix. It can be treated in several ways. One of them consists in calculating the integral representation for logarithm of a matrix. The method was developed in [49], where it was demonstrated in details how the full infinite set of solutions can be found. It is based on the classical theorem of matrix theory, stating that any nonsingular (real or complex) square matrix $\hat{U}$ possesses a logarithm, i.e. there exists a matrix $\hat{A}$, such that $\hat{U} = \exp \hat{A}$. In particular, the theorem was proved, stating that for a nonsingular matrix $\hat{U}$ and for an angle $\theta$, such that $e^{i\theta}\hat{U}$ has no singular values on $(-\infty, 0]$, the equation $\hat{U} = \exp \hat{A}$ has the solution

$$\hat{A} = \log \hat{U} = \int_0^1 \left[e^{i\theta}\hat{U} - I\right]\left[(1-t)I + te^{i\theta}\hat{U}\right]^{-1} dt - i\theta I . \tag{14}$$

This logarithmic solution is an analytic matrix function and it commutes with any matrix, which commutes with $\hat{A}$. Moreover, any solution that commutes with $\hat{A}$ differs from the above one by a logarithm of the unity matrix (i.e., a solution of $I = \exp \hat{A}$. Based on the above proved statement [49], we can proceed on the simplest supposition of $\theta = 0$, which yields

$$\hat{A} = \log \hat{U} = \int_0^1 \left[\hat{U} - I\right]\left[t(\hat{U} - I) + I\right]^{-1} dt . \tag{15}$$

Other values of $\theta$ are possible, but the above simplest form is good for our calculations of the exponential mixing matrix.

The other, not much more simple, but pure algebraic method to calculate the matrix logarithm consists in the use of the Jordan form $\hat{J}_U = \hat{E}^{-1}\hat{U}\hat{E} = \mathrm{diag}(\varepsilon_1, \varepsilon_2, \varepsilon_3)$ of the matrix $\hat{U}$, where $\hat{E} = (e_1, e_2, e_3)$, $e_i$ are eigenvectors and $\varepsilon_i$ are eigenvalues of the proper equation: $\hat{U}e_i = \varepsilon_i e_i$. Then for the function $f(\hat{U})$ we have the Jordan form $\hat{J}_{f(\hat{U})} = \hat{E}^{-1} f(\hat{U})\hat{E} = \mathrm{diag}(f(\varepsilon_1), f(\varepsilon_2), f(\varepsilon_3))$. It is now easy to obtain the desired presentation for $f(\hat{U}) = \hat{E}\hat{J}_{f(\hat{U})}\hat{E}^{-1}$. Thus for the specific case of the matrix logarithm we have

$$\log \hat{U} = \hat{E}\, \mathrm{diag}(\log(\varepsilon_1), \log(\varepsilon_2), \log(\varepsilon_3))\hat{E}^{-1}, \tag{16}$$

where $\varepsilon_i$ are eigenvalues and $e_i$ are eigenvectors of the equation $\hat{U}e_i = \varepsilon_i e_i$, composing the matrix $\hat{E} = (e_1, e_2, e_3)$. Recent application of this technique for Hamiltonian operators was done in [50].

The exponential parameterization of the mixing matrix allows factorization of the rotational and of the CP-violating parts [36] as follows:

$$\mathbf{U} = \mathbf{R}\mathbf{P}_{CP}.\tag{17}$$

The rotation part is given by the real exponential matrix in the angle–axis form:

$$\mathbf{R} \equiv \begin{pmatrix} R_{xx} & R_{xy} & R_{xz} \\ R_{yx} & R_{yy} & R_{yz} \\ R_{zx} & R_{zy} & R_{zz} \end{pmatrix} = e^{\mathbf{A}_{\text{Rot}}} = e^{\begin{pmatrix} 0 & \lambda & \mu \\ -\lambda & 0 & \nu \\ -\mu & -\nu & 0 \end{pmatrix}} = e^{\Phi \begin{pmatrix} 0 & -n_z & n_y \\ n_z & 0 & -n_x \\ -n_y & n_x & 0 \end{pmatrix}}, \tag{18}$$

which represents the generator of the rotation in three dimensions in the angle $\Phi$ around the axis, defined by the vector $\vec{\mathbf{n}} = (n_x, n_y, n_z)$. In this case $\phi = 0 = \det \mathbf{H}$ and (9) reduces to the well known Rodrigues formula for $SO(3)$ rotations about the axis $\vec{\mathbf{n}}$:

$$\exp[i\theta\mathbf{H}]_{\phi=0} = \mathbf{I} + i\mathbf{H}\sin\theta + \mathbf{H}^2(\cos\theta - 1), \tag{19}$$

thus providing the embedding $SO(3) \subset SU(3)$. The entries $R_{ij}$ of the rotation matrix (18) are expressed in terms of the angle of real rotation $\Phi$ and the vector $\vec{\mathbf{n}}$ as follows [51]:

$$R_{ij} = (1 - \cos\Phi)n_i n_j + \delta_{ij}\cos\Phi - \varepsilon_{ijk}n_k \sin\Phi, \; i, j, k = x, y, z, \tag{20}$$

where $\delta_{ij}$ is the Kronecker symbol, $\varepsilon_{ijk}$ is the Levi-Civita symbol. The angle $\Phi$ in (18) is composed of the entries of the exponential matrix $\mathbf{A}_{\text{Rot}}$ as follows:

$$\Phi = \pm\sqrt{\lambda^2 + \mu^2 + \nu^2}, \tag{21}$$

and the coordinates of the axis $\vec{\mathbf{n}} = (n_x, n_y, n_z)$ are expressed via the parameters $\lambda, \mu, \nu$:

$$n_x = -\frac{\nu}{\Phi}, n_y = \frac{\mu}{\Phi}, n_z = -\frac{\lambda}{\Phi}. \tag{22}$$

Thus, when the CP violation is absent, $\delta = 0$, we end up with the above described real space rotation $\mathbf{R} = e^{\mathbf{A}_{\text{Rot}}}$. In the presence of the CP violation we can separate the real and the imaginary parts of the matrix $\mathbf{A}$ in the exponential parameterization $\mathbf{U}_{\exp} = \exp \mathbf{A}$ (7):

$$\mathbf{A}_{\text{Rot}} = \text{Re}[\mathbf{A}], \; \mathbf{A}_{CP} = i\,\text{Im}[\mathbf{A}]. \tag{23}$$

Then the CP violation is accounted for by the exponential matrix

$$\mathbf{P}_{CP} = \exp(\mathbf{A}_{CP}). \tag{24}$$

Moreover, we can rewrite generic exponential parameterization (7) with the help of the well known formula from the theory of matrices in the following form:

$$\mathbf{U}_{exp} = e^{\hat{\mathbf{A}}} = \exp(\hat{\mathbf{A}}_1 + \hat{\mathbf{A}}_2) = \exp\left(\frac{\hat{\mathbf{A}}_1}{2}\right)\exp(\hat{\mathbf{A}}_2)\exp\left(\frac{\hat{\mathbf{A}}_1}{2}\right) + \frac{1}{4!}\left[\hat{\mathbf{A}}_1\left[\hat{\mathbf{A}}_1,\hat{\mathbf{A}}_2\right]\right] + \ldots \quad (25)$$

The relation (25) for the exponential matrix yields in fact the new parameterisation, which involves the rotation matrix $\mathbf{P}_{Rot}$ and the CP-violating matrix:

$$\tilde{\mathbf{U}} = \mathbf{P}_{CP/2}\mathbf{P}_{Rot}\mathbf{P}_{CP/2}. \quad (26)$$

Matrix $\mathbf{P}_{CP/2}$, accounting for the imaginary term contribution, i.e. for the CP violating part, reads as follows:

$$\mathbf{P}_{CP/2} = \exp\left(\frac{\mathbf{A}_{CP}}{2}\right). \quad (27)$$

Direct check of the unitarity of the matrix $\tilde{\mathbf{U}}$ (26) confirms that the new parameterization $\tilde{\mathbf{U}} = \mathbf{P}_{CP/2}\mathbf{P}_{Rot}\mathbf{P}_{CP/2}$ is exactly unitary.

### 3. Real rotation matrix and the current experimental data

Experimental values for the mixing angles of neutrinos [14], [29], are less well determined than those for quarks; according to the most recent data [52] (May 2016), the average values of these angles for neutrinos read as follows:

$$\theta_{12} \cong 33.72°, \quad \theta_{23} \cong 49.3°, \quad \theta_{13} \cong 8.47°, \quad \delta = 272°. \quad (28)$$

The above values are quite close to those of the TBM parameterization, but for $\theta_{13}$, which is small, but not zero. The best fit, based upon the above given mixing angles, gives the following mixing matrix:

$$\mathbf{U}_{best\ fit} = \begin{pmatrix} 0.823 & 0.549 & 0.005 + 0.147i \\ -0.365 + 0.093i & 0.540 + 0.062i & 0.750 \\ 0.418 + 0.080i & -0.632 + 0.053i & 0.645 \end{pmatrix}. \quad (29)$$

The absolute values of this mixing matrix read as follows:

$$|\mathbf{U}_{best\ fit}| = \begin{pmatrix} 0.823 & 0.549 & 0.147 \\ 0.374 & 0.546 & 0.750 \\ 0.428 & 0.633 & 0.645 \end{pmatrix}. \quad (30)$$

Without the CP violation for $\delta = 0$ we obtain:

$$\left|\mathbf{U}_{st}\right|_{\delta=0} = \begin{pmatrix} 0.823 & 0.549 & 0.147 \\ \mathbf{0.455} & \mathit{0.480} & 0.750 \\ \mathit{0.341} & \mathbf{0.684} & 0.645 \end{pmatrix}. \tag{31}$$

Upon the comparison of (31) with (30) we see that the bold values (2,1), (3,2) of the $\left|\mathbf{U}_{st}\right|$ for $\delta = 0$ (see (31)) are greater and the italicised values (2,2), (3,1) are smaller than those calculated for $\delta = 272°$ (see (30)). The first line of the matrix is best determined, as well as the element (3,3); other values are in relatively broad range. As regards the CP violation, there are only indications that the CP-violating phase value is $\delta \approx 272°$, this value being quite approximate (see, [29], [53]). Note, that the entry (1,3) is definitely not equal zero, contrary to that in the TBM parameterization.

First of all we will explore exponential parameterization of the neutrino mixing matrix with the current data set (29) and compare with that, resulting from the TBM form. We will omit the Majorana phases for simplicity; it was shown, that they interplay with the CP phase in some entries of the exponential matrix and just produce more complex terms (see [36]). Let us consider first of all only the real rotational part and compare the rotational matrix (18) with the TBM form of the mixing matrix [21], [29]. With the help of (19), (20), (21), (22), we obtain the following values for the parameters of the exponential parameterisation (18), corresponding the TBM parameterization:

$$\lambda_{TBM} \cong 0.5831, \ \mu_{TBM} \cong -0.2415, \ \nu_{TBM} \cong 0.7599. \tag{32}$$

This yields the following coordinates of the rotational axis and angle:

$$\vec{\mathbf{n}}_{TBM} = (0.7858, 0.2235, 0.5777), \ \Phi_{TBM} \cong 56.6°. \tag{33}$$

Now, from the data set, reported in [14], [29], we obtain for neutrinos

$$\lambda_{2014} \cong 0.516, \ \mu_{2014} \cong -0.342, \ \nu_{2014} \cong 0.611; \tag{34}$$

the corresponding rotational axis and angle read as follows:

$$\vec{\mathbf{n}}_{\nu 2014} = (0.7021, 0.3936, 0.5934), \ \Phi_{\nu 2014} \cong 49.8°. \tag{35}$$

To avoid the uncertainty, originating from largely undetermined CP-violating phase, we calculate the fit with the experimentally determined values of the entries of the PMNS matrix, which contain only the mixing angles $\theta_{i,j}$ and do not depend on the CP violation, described by $\delta$. We obtain the following values for the entries of the rotational matrix $\mathbf{R}$:

$$\lambda_{2016} \cong 0.61396, \ \mu_{2016} \cong -0.11845, \ \nu_{2016} \cong 0.87681, \tag{36}$$

which yield the axis–angle rotation with

$$\vec{n}_\nu = (0.8142,\ 0.110,\ 0.5701),\quad \Phi_\nu \cong 61.7°. \tag{37}$$

All the above results are obtained for the unitary exponential mixing matrix and are based either on the unitary TBM matrix (32), (33) or on two different data sets with respective best fit mixing angles; the precision is determined exclusively by the errors in the experimental data evaluation and fit. The real rotational matrix $\mathbf{A}_{\text{Rot}}$ with (36), (37) yields the mixing matrix $\mathbf{U}_{\text{exp}}$ values, which coincide with the $|\mathbf{U}_{\text{st}}|$ matrix values for $\delta=0$ (31).

Comparing the coordinates of the rotation vectors and the angles, obtained from the TBM parameterization (33) [29], from the data of 2014-2015 (35) [14], [29] and from the most recent data (May 2016) (37) [52], we see that the resulting mixing matrix parameters $\lambda$, $\mu$, $\nu$, (see (32), (34), (36)) and the respective angle–axis rotations (33), (35), (37) differ from each other quite much. For example, the value of the entry $\mu$ has decreased from $\mu \approx 0.3$ in the year 2014 to $\mu \approx 0.1$ in May 2016. The value of the "$n_y$" coordinate of the rotation vector varies in wide range [0.1–0.4] dependently on the data set and the year. Surprisingly, the coordinates of the rotation vector and the angle (37) based on the most recent data set (May 2016) [52] are closer to the results based on the exactly unitary TBM values, than the ~1.5 year old data [14], [29] based result (35) is. The same observation regards the rotation angle, which varies $\approx 20\%$ from $\approx 50°$ to $\approx 62°$, dependently on the data set. However, with all the above differences in the rotation coordinates obtained from different data sets in different years, the angle between the rotation axes of quarks and the rotation axes of neutrinos remains remarkably stable $\approx 45°$. Indeed, taken the well determined small values of the mixing angles for quarks $\theta_{q12}=13.04°$, $\theta_{q23}=2.38°$, $\theta_{q13}=0.201°$, we obtain the direction of the rotation vector (22) for quarks:

$$\vec{n}_q = (0.1829, 0.0206, 0.9831). \tag{38}$$

Comparing it with the above determined coordinates of the rotation vectors for neutrinos TBM (33), years 2014-15 (35) and May 2016 (37), we obtain respectively 43.6°, 44.5° and 45.8° degrees.

Thus, while the experimental data on the neutrino mixing changes from year to year and from set to set with some 20% and for some values even more, the QLC hypothesis

[31], [22], which states that the angle between the rotation axes of quarks $\vec{\mathbf{n}}_q$ and of neutrinos $\vec{\mathbf{n}}_\nu$ constitute the 45° angle, holds well, because its value varies with just about 1%. This is much lower than the error margins of the experimental data sets. This hypothesis, however, still does not have sufficient physical reasons or theoretical fundaments.

### 4. Exponential mixing matrix with account for the CP violation

So far the account for the CP violation in the exponential parameterization [36] has been conducted via the same scheme as for quarks, using eq.n (8). Even recently it has lead to satisfactory results in description of the CP violation [54]. However, with the most recent data on the CP violation in the lepton sector [52] this approach yields the entries for the unitary mixing matrix, which deviate far from the experimental values. To overcome this difficulty of the exponential parameterization with complex exponents in just (1,3) and (3,1) entries, we develop in what follows a precise account for the CP violation by means of the matrix logarithm technique, described in section 2. Current estimations of the CP violation are based on indirect experimental observations and they remain largely approximate. Nevertheless we will consider the present best fit matrix (29) and its absolute values (30) as reference data set. The fundamental problem of constructing the exponential mixing matrix (7), giving exactly the best fit (29), can be solved in several ways, in particular with the help of the integrals (14), (15) or by alternative method, using the matrix Jordan form and eq. (16). Both of these methods work well and straightforward calculations of the formulae (15) or (16) yield the identical result for the exponential matrix $\mathbf{A}$, reproducing the best fit (29):

$$\mathbf{A} = \begin{pmatrix} -0.0253632i & 0.551703 + 0.0557619i & -0.249131 + 0.136429i \\ -0.551703 + 0.0557619i & 0.0502214i & 0.834211 + 0.0319945i \\ 0.249131 + 0.136429i & -0.834211 + 0.0319945i & -0.0248583i \end{pmatrix}. \quad (39)$$

Thus we have obtained the matrix logarithm $\mathbf{A}$ (39) of the best fit matrix (29). It is worth saying that the diagonal entries of (39) are not zeroes, $\mathbf{A}$ is traceless: $\text{Tr}[\mathbf{A}] = 0$, and it has anti Hermitian form, which ensures the unitarity of the proper transform. With account of the best fit values (29) and using (12) and (13), we obtain for the exponential matrix as

the SU(3) group element the following values of the SU(3) group rotation angle $\theta$ and the angle $\phi$:

$$\theta = 1.0426 = 59.74°, \ \phi = 0.105501 = 6.05°, \tag{40}$$

which cyclometrically encodes the invariant

$$\det \mathbf{H} = -\frac{2}{3\sqrt{3}} \sin 3\phi = -0.119799. \tag{41}$$

Since $\mathbf{A} = i\theta\mathbf{H}$, we obtain the following entries of the matrix $\mathbf{H}$:

$$\mathbf{H} = \begin{pmatrix} -0.024327 & 0.053484 - 0.529167i & 0.130856 + 0.238954i \\ 0.053484 + 0.529167i & 0.048170 & 0.030688 - 0.800135i \\ 0.130856 - 0.238954i & 0.0306876 + 0.800135i & -0.023843 \end{pmatrix}, \tag{42}$$

and the exponential $\mathbf{U}_{\exp} = \exp[i\theta\mathbf{H}]$ (see (9)) yields exactly the best fit values (29).

The result (9) for the SU(3) element, $\mathbf{U}_{\text{PMNS}}$, generated by the traceless 3×3 Hermitian matrix $\mathbf{H}$, appears in terms of elementary triginometric functions since the invariant $\det \mathbf{H}$ is expressed as the angle $\phi$. This allows to obtain analytical expressions for all the entries of the exponential matrix $\mathbf{A}$ as functions of the angles of the standard parameterization $\mathbf{U}_{\text{st}}$ (3). However these expressions are huge and cumbersome, they do not bring any more clarity and we omit them for conciseness.

Differently from the previously used simplest account for the CP violation by the exponential of $\mathbf{A}_0$ (8), where only the entries (1,3) and (3,1) were complex and which does not exactly match the best fit, the obtained matrix $\mathbf{A}$ (39) contains imaginary diagonal entries and all the other entries are complex:

$$\mathbf{A} = \mathbf{A}_{\text{Rot}} + \mathbf{A}_{\text{CP\_1}} + \mathbf{A}_{\text{diag Im}}. \tag{43}$$

We separated the real part of the matrix $\mathbf{A}$:

$$\mathbf{A}_{\text{Rot}} = \text{Re}[\mathbf{A}] = \begin{pmatrix} 0 & 0.551703 & -0.249131 \\ -0.551703 & 0 & 0.834211 \\ 0.249131 & -0.834211 & 0 \end{pmatrix}, \tag{44}$$

which gives the real space rotation $\mathbf{R} = e^{\mathbf{A}_{\text{Rot}}}$, the diagonal imaginary elements of $\mathbf{A}$ in the form of the matrix

$$\mathbf{A}_{\text{diag Im}} = i\,\text{diag}\{\alpha_1, \alpha_2, \alpha_3\},$$
$$\alpha_1 = -0.0253632, \alpha_2 = 0.0502214, \alpha_3 = -0.0248583, \tag{45}$$

whose entries sum equals zero: $\alpha_1 + \alpha_2 + \alpha_3 = 0$, and the imaginary part $\mathbf{A}_{CP\_1}$ of the non-diagonal entries of $\mathbf{A}$, which provides the major account for the CP violation:

$$\mathbf{A}_{CP\_1} = i\,\mathrm{Im}[\mathbf{A} - \mathbf{A}_{\text{diag Im}}], \tag{46}$$

$$\mathbf{A}_{CP\_1} = \begin{pmatrix} 0 & 0.0557619i & 0.136429i \\ 0.0557619i & 0 & 0.0319945i \\ 0.136429i & 0.0319945i & 0 \end{pmatrix}. \tag{47}$$

Note, that all the entries in $\mathbf{A}_{CP\_1}$ are complex and not only (1,3), (3,1) entries as in $\mathbf{A}_0$, used for CP violation account in [36], [54]. Apart from the minor diagonal imaginary elements $\mathbf{A}_{\text{diagIm}}$, the matrix $\mathbf{A}$ accounts for the CP violation in the form of rotation (18) around the axis, whose coordinates have complex values $\lambda_i$, dependent on the CP phase $\delta$:

$$\mathbf{A} = \begin{pmatrix} 0 & \lambda_1(\delta) & \lambda_3(\delta) \\ -\lambda_1^*(\delta) & 0 & -\lambda_2(\delta) \\ -\lambda_3^*(\delta) & \lambda_2^*(\delta) & 0 \end{pmatrix} + \mathbf{A}_{\text{diag Im}}. \tag{48}$$

Interestingly, the pure imaginary diagonal exponential term $\mathbf{A}_{\text{diagIm}}$ produces the diagonal exponential matrix with imaginary entries

$$\mathbf{P}_{\text{diag Im}} = \mathrm{diag}\{e^{i\alpha_1}, e^{i\alpha_2}, e^{i\alpha_3}\}, \tag{49}$$

which reminds the Majorana term. However, the matrix $\mathbf{P}_{\text{diag Im}}$ originates from the CP violation. The respective phases are very small $\sim 10^{-2}$ (see (45)).

For the Jordan form (16) we obtain the following eigenvectors

$$e_1 = \begin{pmatrix} 0.812 \\ 0.231 + 0.045i \\ 0.531 - 0.053i \end{pmatrix}, \quad e_2 = \begin{pmatrix} 0.358 \\ -0.236 + 0.683i \\ -0.529 - 0.265i \end{pmatrix}, \quad e_3 = \begin{pmatrix} 0.461 \\ -0.225 - 0.610i \\ -0.524 + 0.299i \end{pmatrix}. \tag{50}$$

They compose the matrix $\hat{E} = (e_1, e_2, e_3)$ and together with the eigenvalues:

$$\varepsilon_1 = 0.992 + 0.126i, \quad \varepsilon_2 = 0.562 + 0.827i, \quad \varepsilon_3 = 0.454 - 0.891i \tag{51}$$

they yield the same result (39) for the matrix $\mathbf{A}$. Direct substitution of (39) into (7) yields the best fit matrix (29). Thus we have obtained the exponential parameterization, exactly reproducing the best fit: $e^{\mathbf{A}} = \mathbf{U}_{\text{best fit}}$. Moreover, with the Jordan form we employ purely algebraic method for the solution of the characteristic equation and the consequent substitution in (16) allows explicit analytical relation between the entries of the mixing matrix **U** and the exponential matrix **A**. However, they appear cumbersome and very bulky, so we omit them for brevity. Numerical calculations of (15) or (16), yielding the results (51), (50), (39), are straightforward and simple. Now that we have obtained the exact exponential parameterization (7) for the best fit by constructing the matrix logarithm, we can evaluate the influence of the entries of the obtained matrix **A** on the neutrino mixing. Upon the comparison of the entries of (39) with each other it is easy to recognize that the major contribution of the CP violation comes in the entries (1,3) and (3,1). The nonzero complexity of other non-diagonal entries plays secondary role and the diagonal elements of the exponential **A**, which are imaginary, play even minor role since their absolute value is very small as compared with that of all other entries. Indeed, expressed in terms of the angles, the entries of **A** read as follows:

$$\mathbf{A} = \begin{pmatrix} -0.0253632\,e^{i90°} & 0.554514\,e^{i5.7°} & 0.284041\,e^{i151.3°} \\ -0.554514\,e^{i(180°-5.7°)} & 0.0502214\,e^{i90°} & 0.834825\,e^{i2.2°} \\ -0.284041\,e^{i(180°-151.3°)} & -0.834825\,e^{i(180°-2.2°)} & -0.0248583\,e^{i90°} \end{pmatrix}. \quad (52)$$

This form evidences that the major complex contribution due to the CP violation is in the corner entries (1,3) and (3,1) as proposed in (8). These entries regard only electron and taon neutrinos with the CP violating angle $\approx 150°$ angle and $\approx 0.28$ absolute value. The CP angles for other neutrino pairs are much smaller: few degrees only. It also confirms the validity of the previously developed in [36] approximate approach, which accounts for the complexity only in the entries (1,3) and (3,1) (see (8)) and thus yields the imaginary CP violating matrix $\mathbf{A}_{CP}$ (see (23)) in the following form:

$$\mathbf{A}_{CP\_0} = \begin{pmatrix} 0 & 0 & 0.136429\,i \\ 0 & 0 & 0 \\ 0.136429\,i & 0 & 0 \end{pmatrix}. \quad (53)$$

The respective exponentials of the CP-violating matrices $\mathbf{A}_{CP}$, $\mathbf{A}_{CP\_1}$ and $\mathbf{A}_{CP\_0}$ can be expressed as matrix polynomials second-order in **H** with the help of (19). Based on the best fit data (29) with account for (53) we obtain for the entries of $\mathbf{P}_{CP}$ matrix (24) the following values:

$$\mathbf{P}_{CP\_0} = \begin{pmatrix} 0.990708 & 0 & 0.136006i \\ 0 & 0 & 0 \\ 0.136006i & 0 & 0.990708 \end{pmatrix}. \qquad (54)$$

Note, that $\mathbf{A}_{CP\_0}$ (53) is in fact the matrix (8) for the CP angle $\delta = 151.3°$. More accurate account for the CP violation can be done with the help of the matrix $\mathbf{A}_{CP\_1}$ (47) with complex nonzero values for all non-diagonal entries, which yields the matrix

$$\mathbf{P}_{CP\_1} = \begin{pmatrix} 0.989159 - 0.000081i & -0.002177 + 0.055551i & 0.135912i \\ -0.002177 + 0.055551i & 0.997937 - 0.000081i & -0.003796 + 0.031873i \\ 0.135912i & -0.003796 + 0.031873i & 0.9902 - 0.000081i \end{pmatrix}. \quad (55)$$

The transform by this matrix involves all three neutrino types, while the transform by $\mathbf{P}_{CP\_0}$ concerned only electron and taon neutrinos.

The closest approximation is given by the exponential with the CP-violating matrix, including the diagonal elements:

$$\mathbf{A}_{CP} = i\,\mathrm{Im}[\mathbf{A}] = \begin{pmatrix} -0.0253632i & 0.0557619i & 0.136429i \\ 0.0557619i & 0.0502214i & 0.0319945i \\ 0.136429i & 0.0319945i & -0.0248583i \end{pmatrix}, \qquad (56)$$

and the respective factorization involves the matrix

$$\mathbf{P}_{CP} = \begin{pmatrix} 0.98884 - 0.025207i & -0.002870 + 0.055533i & 0.002527 + 0.13587i \\ -0.002870 + 0.055533i & 0.996679 + 0.050068i & -0.004201 + 0.031863i \\ 0.002527 + 0.13587i & -0.004201 + 0.031863i & 0.989894 - 0.024704i \end{pmatrix}. \quad (57)$$

From the comparison of (56) with (47) and of (57) with (55) we see that the values of the entries of the matrices $\mathbf{P}_{CP}$, $\mathbf{P}_{CP\_1}$ do not differ from each other that much. We conclude that the CP violation can be viewed as the rotation in the imaginary space around the vector with imaginary coordinates, complemented by the exponential with the imaginary diagonal matrix $\mathbf{A}_{\mathrm{diagIm}}$, whose form resembles that of the Majorana term.

Now we can compare the resulting exponential neutrino mixing matrix (7) with the factorization (17) of the contributions of the real three dimensional space rotation $\mathbf{A}_{\mathrm{Rot}}$ and of the CP violation, accounted by one of the matrices $\mathbf{A}_{CP}$, $\mathbf{A}_{CP\_1}$ or $\mathbf{A}_{CP\_0}$. Evidently, (7) with account for (39) gives exactly the best fit matrix (29):

$$\mathbf{U}_{\mathrm{Rot+CP}} = \exp[\mathbf{A}_{\mathrm{Rot}} + \mathbf{A}_{CP}] = \mathbf{U}_{\mathrm{best\,fit}}. \qquad (58)$$

If we neglect the minor diagonal terms (45) and substitute the matrix $\mathbf{A}_{\text{Rot}} + \mathbf{A}_{\text{CP\_1}}$ in (7), then we obtain the following absolute values for the neutrino mixing matrix:

$$\left|\mathbf{U}_{\text{Rot+CP\_1}}\right| = \left|\exp[\mathbf{A}_{\text{Rot}} + \mathbf{A}_{\text{CP\_1}}]\right| = \begin{pmatrix} 0.823 & 0.550 & 0.141 \\ 0.378 & 0.542 & 0.751 \\ 0.424 & 0.635 & 0.646 \end{pmatrix}, \tag{59}$$

which is in very good agreement with the best fit values (30). Thus, the contribution of the CP originated imaginary diagonal matrix $\mathbf{A}_{\text{diagIm}}$ with very small angles, which has the form of the Majorana term, is, in fact, just the fine tune to the major contribution of the $\mathbf{A}_{\text{CP\_1}}$ matrix (47) in the CP violation description. The ansatz $\mathbf{U}_{\text{Rot+CP\_1}}$, which constitutes rotations around axes with real and imaginary coordinates and employs the exponential parameterization matrix with zero diagonal entries, while being a good approximation, does not exactly reproduce the best fit matrix values (29) and (30).

The simplest account for the major CP violation term concerns just two neutrino types: electron and taon. It is done with the help of the matrix $\mathbf{A}_{\text{CP\_0}}$ and it gives quite good agreement with the best fit, but for the entry (1,3), which is too small:

$$\left|\mathbf{U}_{\text{Rot+CP\_0}}\right| = \left|\exp[\mathbf{A}_{\text{Rot}} + \mathbf{A}_{\text{CP\_0}}]\right| = \begin{pmatrix} 0.824 & 0.555 & \mathit{0.112} \\ 0.366 & 0.543 & 0.755 \\ 0.431 & 0.630 & 0.646 \end{pmatrix}. \tag{60}$$

The factorization of the real space rotation and the CP violation by account for the leading term $\mathbf{A}_{\text{CP\_0}}$ (53) yields the following mixing matrix:

$$\left|\mathbf{U}\right| = \left|\mathbf{RP}_{\text{CP\_0}}\right| = \begin{pmatrix} 0.825 & 0.554 & \mathit{0.113} \\ 0.375 & 0.543 & 0.752 \\ 0.432 & 0.631 & 0.650 \end{pmatrix}, \tag{61}$$

which, apart the entry (1,3) is also close to the best fit (30). If we factorize the whole imaginary part of the exponential matrix $\mathbf{A}$ (39), then we end with the matrix

$$\left|\mathbf{U}\right| = \left|\mathbf{RP}_{\text{CP}}\right| = \begin{pmatrix} 0.822 & 0.555 & 0.131 \\ 0.387 & 0.540 & 0.748 \\ 0.419 & 0.633 & 0.651 \end{pmatrix}, \tag{62}$$

which is in good agreement with the best fit values (30), and only the entry (1,3) is little bit lower than that in the best fit matrix. The factorization with the imaginary rotational exponential matrix $\mathbf{A}_{CP\_1}$: $\mathbf{U} = \mathbf{RP}_{CP\_1}$, gives practically the same results as $|\mathbf{U}| = |\mathbf{RP}_{CP}|$ (62):

$$|\mathbf{U}| = |\mathbf{RP}_{CP\_1}| = \begin{pmatrix} 0.822 & 0.553 & 0.131 \\ 0.386 & 0.539 & 0.749 \\ 0.417 & 0.635 & 0.650 \end{pmatrix}. \qquad (63)$$

Inclusion of the diagonal imaginary term as a factor: $\mathbf{U} = \mathbf{RP}_{CP\_1} \mathbf{P}_{diag\,Im}$, expectedly changes the phases of the entries slightly and does not effect the absolute values of $\mathbf{U}$. The factorization by $\tilde{\mathbf{U}} = \mathbf{P}_{CP/2} \mathbf{P}_{Rot} \mathbf{P}_{CP/2}$ yields the results similar to those above. Thus, we have demonstrated a number of possible exponential factorizations of the neutrino mixing matrix, which are in good agreement with the best fit. The unitarity of the PMNS transform is ensured by the anti-Hermitian form of the CP-violating matrices.

## Conclusions

The exponential parameterization of the mixing matrix for the neutrinos is applied for the comparative analysis of the mixing data from tri-bimaximal parameterization, data of 2014 year and May 2016 year, this latter with account for the CP violation. The analysis of the mixing matrix values without the CP angle $\delta$ shows that the angle of the rotation in the real three dimensional space varies from one data set to another in the range from 50° to 62°. We have calculated proper entries of the exponential mixing matrix for all studied data sets; they are given by formulae (32), (34), (36). The direction of the rotation axis in space also changes, dependently on which data set we consider. Proper rotation vectors and angles for different sets are $\vec{\mathbf{n}}_{TBM} = (0.7858, 0.2235, 0.5777)$, $\Phi_{TBM} \cong 56.6°$, $\vec{\mathbf{n}}_{\nu 2014} = (0.7021, 0.3936, 0.5934)$, $\Phi_{\nu 2014} \cong 49.8°$, $\vec{\mathbf{n}}_{\nu 2016} = (0.8142, 0.110, 0.5701)$, $\Phi_{\nu 2016} \cong 61.7°$. The result, based on the May 2016 data, appears closer to the rotation vector of the tri-bimaximal parameterization, than that, based on 2014 year data. Despite relatively large spread in the coordinates of the rotation vectors and in the rotation angles, the angle between the axes for quarks and for neutrinos remains $\cong 45° \pm 1°$ degrees, and this confirms the hypothesis of complementarity for neutrinos and quarks [22], [31].

The $U_{PMNS}$ matrix as the element of the SU(3) group is considered. The effect of the CP violation in the framework of the exponential mixing matrix is explored. The value of the group rotation angle $\theta$ is found: $\theta = 1.0426 = 59.74°$; the other angle $\phi = 0.105501 = 6.05°$ encodes the invariant $\det H$. The exponential parameterization (7) $U_{exp} = \exp[A] = \exp[i\theta H]$ is expressed as a matrix polynomial of the second order, where the group element depends on the sole invariant $\det H$ and on the group rotation angle $\theta$. Both dependencies are in terms of elementary trigonometric functions, because $\det H$ is expressed as the angle $\phi$. The exact values of the entries of the Hermitian 3×3 matrix $H$ are obtained in (42). The exponentials of the CP-violating matrices $A_{CP}$, $A_{CP\_1}$, $A_{CP\_0}$ and of the real rotation matrix $A_{Rot}$ can be expressed as second-order matrix polynomials of $H$ by the Euler-Rodrigues result (19).

The logarithm of the mixing matrix is constructed and thus the exact match of the exponential parameterization with the best fit matrix is obtained. It establishes the relation between the proper entries and, in particular, we obtain for the current best fit CP angle $\delta = 272°$ the proper value of the major CP violating angle in the exponential parameterization equal 151.3°. It determines the leading term in the CP violation matrix, $A_{CP\_0}$, and describes the CP contribution to the mixing of electron and taon neutrinos. Accounting only for this leading term in the exponential matrix, we obtain the neutrino mixing matrix values in good agreement with the best fit data, but for the entry (1,3), whole experimentally determined value is significantly larger. The complete CP matrix $A_{CP}$ includes imaginary diagonal elements $A_{diagIm}$, which remind those of the Majorana particles, but $A_{diagIm}$ has the CP origin and it gives very fine contribution to the main CP violating matrix with zero diagonal entries, $A_{CP\_1}$. The latter well describes the CP violation effect for all three neutrino types and its exponential yields the values, which are in very good agreement with the best fit data set. Evidently, the complete account for all the terms in the exponential, $A_{Rot}+A_{CP\_1}+A_{diag\,Im}$, reproduces the best fit matrix values exactly.

A variety of factorizations are possible in the framework of the exponential parameterization. We have distinguished the factors, corresponding the real rotation in three dimensional space, which describes mixing without CP violation, the rotation around the axis with purely imaginary coordinates, describing the major contribution of the CP violation, and the diagonal imaginary term in the exponential mixing matrix $\mathbf{A}_{\text{diag Im}} = i \operatorname{diag}\{\alpha_1, \alpha_2, \alpha_3\}, \alpha_1 = -0.0253632, \alpha_2 = 0.0502214, \alpha_3 = -0.0248583$, which formally resembles the Majorana phases, but comes due to the CP violation. Interestingly, the following relations appears between these entries: $\alpha_1 \cong \alpha_3 \cong -\alpha_2/2$. Without this term the exact account for the CP violation by the exponential parameterization $\mathbf{U}_{\exp} = \exp[\mathbf{A}]$ is not possible. The commonly known ansatz for the exponential matrix with zero diagonal entries is a good approximation of the mixing matrix $\mathbf{U}_{\text{PMNS}}$.

Exponential parameterization of the mixing matrix and obtained with its help results and interpretations can be useful for the analysis of new experimental data on neutrino oscillations in actual and future experiments.